# Customized Monte Carlo Tree Search for LLVM/Polly's Composable Loop Optimization Transformations

Jaehoon Koo, Prasanna Balaprakash, Michael Kruse, Xingfu Wu, Paul Hovland
Argonne National Laboratory, Lemont, IL 60439
{jkoo,pbalapra,michael.kruse,xingfu.wu,hovland}@anl.gov

Mary Hall
University of Utah, Salt Lake City, UT 84103
mhall@cs.utah.edu

## ABSTRACT

Polly is the LLVM project's polyhedral loop nest optimizer. Recently, user-directed loop transformation pragmas were proposed based on LLVM/Clang and Polly. The search space exposed by the transformation pragmas is a tree, wherein each node represents a specific combination of loop transformations that can be applied to the code resulting from the parent node's loop transformations. We have developed a search algorithm based on Monte Carlo tree search (MCTS) to find the best combination of loop transformations. Our algorithm consists of two phases: exploring loop transformations at different depths of the tree to identify promising regions in the tree search space and exploiting those regions by performing a local search. Moreover, a restart mechanism is used to avoid the MCTS getting trapped in a local solution. The best and worst solutions are transferred from the previous phases of the restarts to leverage the search history. We compare our approach with random, greedy, and breadth-first search methods on PolyBench kernels and ECP proxy applications. Experimental results show that our MCTS algorithm finds pragma combinations with a speedup of 2.3x over Polly's heuristic optimizations on average.

## 1 INTRODUCTION

Most compute-intensive programs, in particular scientific applications in high-performance computing, spend a significant amount of execution time in loops, thus making them the prime target for performance optimizations. Common strategies include replacing them by vendor-provided library calls (such as for BLAS [32] or FFT [14]). However, these are available only for a limited set of kernels, which require the use of a specific data layout in memory and may not be available on all platforms. Therefore, researchers have inevitably sought to optimize the loops in the application itself.

Unfortunately, manual optimization is a time-consuming process that requires intimate knowledge about the target hardware and often results in less maintainable code. Code generators promise solutions by automatically optimizing the synthesized code [18] from a domain-specific problem description but specialized for specific kinds of algorithms, such as LIFT [40] for BLAS and SPIRAL [34]. Such code generators have performance-relevant choices to make that are different for each platform. Often, the choice is determined by using *autotuning* [6]: the process of generating a search space of possible implementations/configurations of a kernel or an application and evaluating a subset of implementations/configurations on a target platform through empirical measurements to identify the high-performing implementation/configuration.

The semantics of a program and its performance optimization can be considered as two separate concerns. With the semantics of an algorithm fixed, the goal of autotuning is to find a variant of the program with the lowest time to solution on a given hardware without changing how the algorithm works. Separation of semantics and optimization parameters is one of the motivations of Halide [37], wherein a kernel and an iteration space are specified in a domain-specific language. This is assigned a separate *schedule* that determines the threads and loop nest that execute the kernel over the entire iteration space. The schedule can be optimized automatically [1, 35, 48] via machine learning (ML) methods.

Another approach is to use directives inside the source code that describe the optimizations to apply, whereas the code itself describes only the semantics. This realizes the separation of concerns on any source code containing loops, without being specific to a domain. In C/C++, pragmas are the most straightforward syntax to convey loop optimization directives [16, 30]. Such pragmas, when added to a loop nest, tell the compiler to apply a specific transformation to the loop, such as unrolling or parallelization.

Polly [22] is the LLVM subproject that adds a polyhedral loop optimizer to LLVM's optimization pipeline. By default, it uses various heuristics to optimize a loop nest, including PLuTo's [8] memory locality optimizer, and generalizes matrix-multiplication recognition and optimization [20], tiling, auto-parallelization, and GPU offloading [23]. Recent work exposed Polly's loop transformations as source code directives to the programmer [29] by parsing pragmas in the source code using Clang [33] and passing them to Polly. These pragmas also build up a search space that can be used for autotuning [31]. These directives are composable; in other words, after applying a transformation, more transformations can be applied to the results of a transformation. The set of additional loop transformations that can be applied depends on the sequence of prior loop transformations. Consequently, the search space for autotuning a sequence of loop transformations is tree-shaped [31] where each node represents a sequence of loop transformations, or *configurations*. The tree search space is potentially unbounded in depth (e.g., the choice of tile size can be any positive integer and another loop transformation can always be added). Consequently, the number of potentially useful configurations is too large to search exhaustively.

Monte Carlo tree search (MCTS) [28], originally proposed to solve games in artificial intelligence, is a promising heuristic method to explore treelike search space. However, out-of-the-box MCTS has several limitations for the treelike search space exposed by Polly's loop transformation pragmas. To address these limitations, we introduce a customized MCTS to explore the configuration space of loop transformation pragmas. The contributions of the paper are threefold:

- We develop an autotuning framework to efficiently explore a dynamically growing treelike search space of the newly developed Clang loop optimization pragmas.
- We design a customized MCTS method with a new reward mechanism, restart strategy, and transfer learning to improve the search efficiency.
- We show that the customized MCTS outperforms Polly's optimization heuristic in 16 out of 24 PolyBench kernels and obtains a speedup of 2.1 on average. On three Exascale Computing Project (ECP) proxy applications, our MCTS surpasses Polly in all three and achieves a speedup of 3.7 on average.

## 2 BACKGROUND

In this section we first describe loop transformation pragmas implemented in LLVM Clang/Polly and our treelike search space that grows dynamically. We then provide the mathematical background and explain the algorithmic mechanism of MCTS.

### 2.1 Loop transformation directives

Loop transformation directives—in our case in the form of pragma annotations in the source code—instruct the compiler to apply a specific optimization to a loop instead of using its own profitability heuristics. If multiple transformations are applied to the same loop, the order in which they are applied is significant. For example, in the following example the loop is first tiled by using a tile size of 32; then the tiling's floor loop (the outer loop) is parallelized.

```
#pragma clang loop parallelize_thread
#pragma clang loop sizes(32)
for (int i = 0; i < n; ++i)
  Body(i);
```

The effect of these directives becomes more evident when we explicitly expand the tiling into loops as shown below. It also illustrates that although the `parallelize_thread` textually appears before the tile directive, it is applied second because the directives apply to what is on the next line.

```
#pragma clang loop parallelize_thread
for (int i1 = 0; i1 < n; i1+=32)              /* floor */
  for (int i2 = i1; i2 < i1+32 && i2 < n; ++i2) /* tile */
    Body(i2);
```

The implementation described in [29] consists of a parser component for LLVM/Clang and a loop transformer with Polly, the polyhedral loop optimizer, as illustrated in Figure 1. It is publicly available in GitHub.[1] Clang parses the pragma directives and attaches them as metadata nodes to the loop represented in LLVM-IR. In LLVM's optimization pipeline, these are seen by Polly. Before applying the directives to the intermediate representation (IR), it first checks whether the code's semantics is preserved and rejects the transformation if it is not preserved. This step can be leveraged within autotuning to check loop transformations that produce incorrect code. The compiler has converted the input into an IR that is intended for easier processing and has implemented analyses required to detect violations to semantic constraints that are not included in the transformation search space.

[1] https://github.com/SOLLVE/llvm-project/tree/pragma-clang-loop

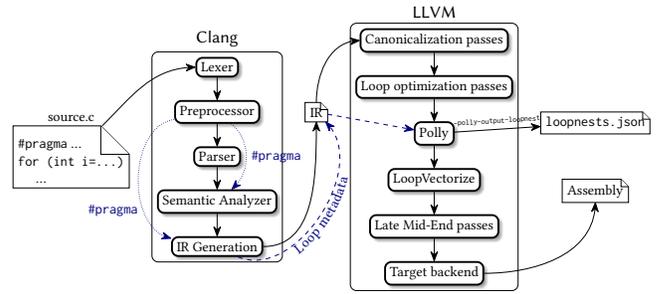

Figure 1: Clang/LLVM compiler architecture

As an alternative to user-directed loop transformations, Polly can use heuristics to optimize a loop nest. The heuristic used by Polly is implemented in ISL [45] using PLuTo [8]. It tries to maximize parallelism and minimize distance. Additionally, Polly applies tiling and vectorization where possible and detects code that implements matrix-matrix multiplication-like algorithms to which it applies a dedicated optimization.

### 2.2 Search space

Most autotuning frameworks and algorithms [3, 5, 38] assume that the tunable parameters can be expressed in a flat list and that the search space is composed of all possible parameter combinations. Nevertheless, the search space of composable loop transformations cannot be expressed in this way. One reason is that the search space is potentially infinite: Any transformation that results in at least one output loop can be applied an arbitrary number of times. This may often not make a lot of sense; for instance, partial unrolling twice with a tile size of four is effectively the same as unrolling with a factor of 16. We consider this domain knowledge for search space pruning. The second reason is that the loop nest structures changes after applying a transformation and therefore the set of choices is changing. As a result, tile sizes from one configuration may be unrelated to the tile sizes of another configuration since they apply to different loops.

In our trials, we use the same search space as in [31] with some additional transformations and other improvements, such as the ability to sparsely instantiate a configuration by a numeric id that does not require instantiating all sibling configurations just to know how many there are. This search space takes the form of a tree, as illustrated in Figure 2. The root node represents the configuration with no loop transformation, in other words, the original program. Every node also represents a loop nest structure to determine which transformations could be applied. The loop nest structure of the root node is determined by Clang, written to a JSON file and processed by the search space generator. Every child node then represents a choice of adding one more transformation, which results in a new configuration.

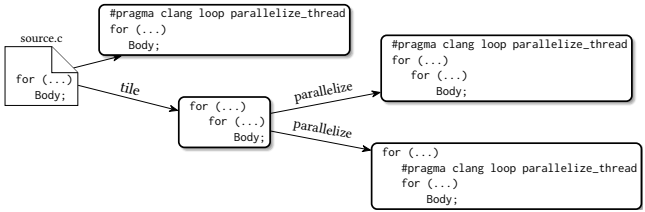

Figure 2: Example of a loop transformation search space. The root represents the original loop without any transformations. It can be transformed by either applying parallelization or tiling, resulting in two new configurations. A parallelized loop is not considered transformable anymore, but tiling results in two new loops that each can be parallelized independently.

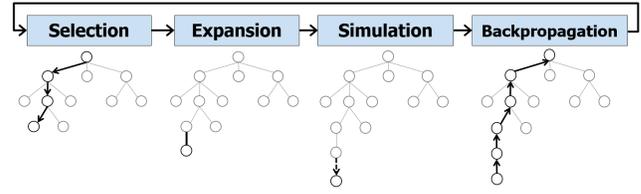

Figure 3: Four phases of MCTS. The selection phase finds a path from the root to the leaf node. The expansion phase creates a child node from the leaf node. The simulation phase conducts simulation from the expanded node until it reaches a terminal node and produces an outcome. In backpropagation, the simulation result is backpropagated through the selected nodes, which is used for the subsequent selection.

Each node/configuration represents a complete program that can be compiled and executed, and its execution time can be measured. The autotuning optimization goal is to find the configuration with the minimal execution time, namely, the best performance. As previously mentioned, however, the compiler may reject a configuration when it cannot ensure that all transformations preserve the program's semantics. In that case, the configuration has to be discarded.

The search space generator applies the following transformations to a node, some of which result in multiple loop transformations with different settings.

- Loop tiling (#pragma clang loop tile): This applies to multiple loops in a perfectly nested loop and applies all possible tile sizes from a preconfigured set. It results in twice as many loops as in the perfect loop nest. The optional clause peel(rectangular) instructs Polly to build a separate loop for the partial tiles, such that the complete tiles do not require conditionals.
- Loop interchange (#pragma clang loop interchange): A new configuration is derived for each permutation of its loops for each perfect loop nest.
- Thread parallelization (parallelize_thread): This parallelizes a single loop using OpenMP, creating the equivalent of #pragma omp parallel for schedule(static). The reason for not using the OpenMP directive directly is that it is not composable with other loop transformations. In Clang/LLVM, OpenMP lowering is implemented in the front end (see Figure 1), whereas loop transformations are implemented by Polly. It is not possible to directly apply #pragma omp parallel for in the mid-end. Furthermore, the OpenMP implementation does not verify the legality of the transformation.
- Loop unrolling (#pragma clang loop unrolling): This involves fully unrolling a loop or partially unrolling by a factor from a predefined set.
- Loop reversal (#pragma clang loop reverse): This executes the iterations of loop in reverse order.
- Array packing (#pragma clang loop pack): Unlike the other directives, this is a data layout transformation. It determines the working set of array elements used in a loop and copies them into a temporary array for the duration of the loop. Polly tries to make the order of the elements in the temporary array resemble the access order. When array elements are also written in the loop, the elements are also written back to the original array. The transformation is most useful either to make the loop's working set fit into a closer cache in the memory hierarchy or to transpose a matrix to make its accesses consecutive within the loop.

## 2.3 Monte Carlo tree search

Monte Carlo tree search (MCTS) is a heuristic search method that seeks to solve a class of computationally intractable sequential decision-making problems, typically represented by trees. In these trees, a node represents a state (decision) in the sequential decision-making process, and the directed edges represent sequential transitions from one state to another. MCTS is an iterative approach that performs four phases at each iteration: selection, expansion, simulation, and backpropagation (see Figure Figure 3 for an illustration). The selection phase traverses only the part of the tree that was already visited. It starts from the root node and selects the next node according to a tree policy. It typically ends in a leaf node of a visited tree. The expansion phase expands the tree by considering children nodes of the ending leaf node of the previous phase and selects one from among them. The simulation phase performs simulations for the new node by visiting subsequent nodes that can be reached from the new node. Typically, this involves several random walks from the new node to various terminal nodes that have outcomes. Based on these outcomes, the reward for each explored random walk is computed starting from the root node. The backpropagation phase propagates the path-specific reward back to the nodes in the given path from the terminal node to the root node. Consequently, the nodes that are common in the high reward branches get reinforced and have a high probability of getting selected in the subsequent iterations.

The *tree policy* used in the selection phase plays a critical role in balancing the exploration of the search space and the exploitation of the already-found promising regions in the search space. A widely used policy is Upper Confidence Bounds applied to Trees (UCT) [4,

28], where a child node $v_j$ is selected to maximize

$$UCT = \bar{R}_j + 2C\sqrt{\frac{2 \ln N}{N_j}}, \quad (1)$$

where $\bar{R}_j$ is the average reward of the node $j$, $N$ is the number of times the current (parent) node has been visited, $N_j$ is the number of times child $j$ has been visited, and $C$ is a constant to balance the exploration and exploitation. The reward range is [0, 1] [10].

## 3 CUSTOMIZED MCTS FOR AUTOTUNING LOOP TRANSFORMATION PRAGMAS

Although MCTS is designed to explore the treelike search space, the vanilla version cannot be applied directly to our autotuning problem. First, there is no concept of playing a game for optimizing loop transformation directives. While the reward is trivial in a game setting with a win or loss, MCTS for autotuning problems requires careful reward engineering. Second, the vanilla MCTS is not efficient if a good solution, a high-performing configuration in our case, is located deep in the tree. Assume that we have a kernel where there are numerous configurations with one directive located at the first depth of the tree. However, the best configuration requires more than one directive. As the pure MCTS expands a node one by one from the root, it requires many iterations to reach the best configuration located deep in the tree. Third, there is no limit on the number of loop transformations to form a configuration (node), which means a terminal node is unknown. In this case the algorithm needs to determine when to stop to form a configuration. Fourth, the vanilla MCTS can end up with a poor local solution.

To circumvent these issues, we developed a customized MCTS for autotuning loop transformation pragmas. Our method consists of two phases: random exploration of configurations at different depths to find promising regions, followed by targeted exploitation of the promising regions. A restart strategy is applied to MCTS to avoid being stuck in a local solution. MCTS repeats the exploration and exploitation after each restart so that configurations at different depths can be searched. Moreover, it leverages the search history after the restart by transferring the high- and low-performing configurations for reinforcement.

### 3.1 Moving average reward function

We design our reward function such that an evaluated configuration (sequence of pragmas applied to the unoptimized code) at each iteration leads to a win and a loss. To this end, we introduce a utility function $h$ that measures the performance of the configuration and a target function $f$ to determine a win or a loss over its performance. We use the utility function that computes the speedup of a configuration $v$ over an unoptimized code located on the root $v_0$. The target is computed by using the $m$-moving average, defined as follows:

$$h_t = \frac{\text{execution time of } v_0}{\text{execution time of } v_t}$$

$$f_t(m) = max(f_{m-1}, \sum_{i=t-m+1}^{t} h_i/m)$$

at iteration $t$, and the reward function is

$$R_t = \begin{cases} R_{\text{penalty}}, & \text{if } \dot{v}_t \\ 1, & \text{if } h_t > f_t \\ 0, & \text{otherwise,} \end{cases} \quad (2)$$

where $\dot{v}_t$ denotes a failure program rejected by the compiler and $R_{\text{penalty}}$ is a negative constant. This reward function progressively increases the ability of MCTS to find configurations whose performances are better than the average performances of the last $m$ configurations.

### 3.2 Random walk to learn depth

Since the depth of the tree is unbounded, MCTS cannot stop at a terminal node in a given iteration. Therefore, learning at what depth to stop is an additional complexity that we have to address in dealing with the autotuning problem.

To that end, we incorporate a random walk search within MCTS to explore configurations at different depths and to find the best depth to terminate the search for each iteration. This is achieved as follows. First, we perform $n$ number of random walks each with randomly generated depth sampled uniformly between 1 and $d_{\max}$. Each random walk is a configuration that corresponds to a sequence of pragmas. We compute their performances by applying them on the unoptimized code and running them on the target hardware. The depth of the configuration with the best performance then is selected as the termination depth for MCTS. We use the random walk not only for learning the depth but also for exploration. The performance values from the evaluation are used to compute the reward and backpropagation.

### 3.3 Restart mechanism to avoid getting stuck in a local solution

Even though MCTS is known to find a relatively good solution, it can get trapped in a local solution [44]. Moreover, the random walk method to learn the depth can exacerbate this issue because the depth once fixed cannot be changed. Specifically, if the random walk did not find a good value for the depth, then MCTS is forced to search with that depth value.

To address these issues, we adopt a restart mechanism, where we detect the convergence to a local solution and restart MCTS with a new random walk. This cycle repeats until a computational budget is exhausted. The convergence of the search is identified if there is no improvement in the target value or the algorithm terminates at the same configuration for more than a threshold.

Since the restart mechanism erases the memory of the MCTS, it cannot leverage what it has seen in the past. To avoid this issue, we transfer information from the search history as soon as a new restart is performed. At each restart, we consider all the configurations evaluated by the MCTS by speedup with respect to the evaluated code. We compute lower and upper quantiles of the speedup distribution given by $\alpha$% and 100-$\alpha$%, where $\alpha$ is a user-defined parameter. For example, when $\alpha$ is set to 5, the lower and upper quantiles are 5% and 95%, respectively. We take configurations with speedup above 100-$\alpha$% and reinforce the path with the positive reward. Similarly, we take the configurations with speedup below $\alpha$% and reinforce the path with the negative reward. It is not guaranteed that all of the pragmas from the lower quantile lead to poor performance. For example, a good pragma can be included in both lower and upper quantiles. Therefore, we penalize the lower quantile configurations that do not share any common pragmas with the configurations that belong to the upper quantile.

**Algorithm 1** Customized MCTS
---
1: Input: $d_{\max}$, budgets for search, convergence settings, and $\alpha$
2: **while** stopping criterion not met **do**
3:     **while** no convergence is detected **do**
4:         $d^* \leftarrow \text{RandomWalk}(d_{\max})$
5:         $\text{Transfer}(\alpha)$
6:         $v^* \leftarrow \text{MCTS}(d^*)$
7:     **end while**
8: **end while**

Algorithm 1 shows the pseudo-code of the customized MCTS method. The outer while loop (lines 2–8) runs until the stopping criterion is met. Typically, this is a wall-clock time or a total number of evaluations allowed. The inner while loop (lines 3–7) comprises a random walk to find the depth $d^*$ (RandomWalk($d_{\max}$)), transfer of configurations from lower and upper quantiles for negative and positive reinforcement (Transfer($\alpha$)), and MCTS with the depth $d^*$ (MCTS($d^*$)). Note that the transfer and the random walk are independent of each other and the order in which they are performed does not matter.

## 4 EXPERIMENTS

We use the PolyBench 4.2 [47] benchmark test suite and ECP proxy applications to evaluate our customized MCTS method. PolyBench consists of 30 numerical computations extracted from various application domains. It covers kernels that include 19 linear algebra computations, 3 image-processing applications, 6 physics simulations, and 2 data-mining processes. We selected 24 kernels with the most levels of nested loops. They are as follows:

- Linear Algebra:
  - BLAS (7): gemm, gemver, gesummv, symm, syr2k, syrk, trmm
  - Kernels (6): 2mm, 3mm, atax, bicg, doitgen, mvt
  - Solvers (3): durbin, lu, ludcmp
- Medley (image processing) (2): floyd-warshall, nussinov
- Physics simulation (stencils) (5): adi, fdtd-2d, jacobi-1d, jacobi-2d, seidel-2d
- Data mining (1): covariance.

We use CoMD, miniAMR, and SW4lite from the ECP proxy applications suite. CoMD[2] is a reference implementation for algorithms used in molecular dynamics; miniAMR[3] implements a basic adaptive mesh refinement method that is often used in physical simulations; and SW4lite[4] is a bare-bones version of the larger SW4 application intended for testing performance optimizations in SW4's most important numerical kernels.

Because of loop distribution, we manually applied loop transformation to increase the number of perfectly nested loops and thus the number of possible transformations such as tiling. That is, we split independent computations in a loop body into two or more separate loops. The kernels from the ECP proxy applications also needed to be customized in order to make them transformable by Polly.

We compared our customized MCTS method with the following search methods.

**Polly compile-time heuristics (O3P):** This is a baseline without autotuning. It uses Clang's highest optimization level -O3 with Polly's optimization heuristic enabled. Additional command line flags are -march=native -mllvm -polly-position=early -mllvm -polly-parallel -mllvm -polly-omp-backend=LLVM -mllvm -polly-scheduling=static that tune the machine code for the target and also enable Polly's auto-parallelizer.

**Random search (RS):** It samples each configuration randomly, wherein a depth is selected at random and a random walk is performed for the selected depth in the search tree.

**Breadth first (BF):** This is the commonly used breadth-first search method.

**Global greedy (GG):** This is a greedy algorithm proposed in [31] for treelike search spaces. For a given node, it evaluates all its children and adds them to a priority queue, starting with the root node. The queue is sorted with the most promising configurations first to be expanded next. We refer the reader to [31] for a detailed exposition of this method.

The experiments were conducted on a Linux machine with 2x Intel Xeon Platinum 8180M CPU @ 2.50 GHz with 384 GB RAM memory; GCC 9.2.0; and Clang 13.0.0 with Polly. The search space includes six loop transformations: loop tiling, loop interchange, thread parallelization, loop unrolling, loop reversal, and array packing. The possible tile sizes are 2, 3, 4, 5, 8, 16, 32, 64, 128, and 256. The choices for unroll factors are 2, 4, and 8. Each configuration was executed five times, and the median of the five was selected as the execution time for the configuration to minimize the impact of noise. We used the execution time measured by PolyBench itself using the -DPOLYBENCH_TIME preprocessor option. This measures only the kernel execution time and excludes overhead such as executable startup time and array initialization. All the search methods except O3P were allowed to run for a wall-clock time of six hours or until 1,000 unique configurations were evaluated. For MCTS, we used an exploration weight $C$ of 0.1 and $\alpha$ of 0.05. Since the first phase does exploration, a relatively small value of exploration weight was selected such that the second phase focused on the exploitation. In the search experiments, $d_{\max}$ is set to five based on a preliminary experiment on depth. Specifically, we found that most of the configurations with the depth value greater than 5 fail to compile or result in code transformation errors. We limit each MCTS run to 300 evaluations. The convergence in the search is detected by lack of improvements for 50 iterations or visiting the same configuration for 10 iterations in a row.

### 4.1 Comparison between search methods

Table 1 shows the comparison of different search methods. For a given kernel, the speedup is given by the ratio of runtimes of the O3P and the best configuration obtained by the search method. We observe that the MCTS obtains better performance than the O3P on 16 out of 24 kernels (speedup values greater than 1). The observed speedups range from 1.01 (durbin) to 8.54 (floyd-warshal). The BF, GG, and RS outperform the O3P only on 8, 11, and 11, kernels, respectively. The average speedups over O3P on 24 kernels are 2.13, 1.42, 1.38, and 1.04 for MCTS, GG, RS, and BF, respectively. Moreover, we observe that MCTS obtains the best speedup values for 10 out of 24 kernels ((boldfaced for each kernel). The GG, BF, and RS methods obtain the best speedup values for 4, 1, and 1 kernels, respectively. On 8 kernels, none of the search methods was able to outperform O3P (italicized in the table).

---
[2]https://github.com/ECP-copa/CoMD
[3]https://github.com/Mantevo/miniAMR
[4]https://github.com/geodynamics/sw4lite

**Table 1: Performance comparison on 24 PolyBench kernels**

| Kernels | Speedup over O3P | | | |
|---|---|---|---|---|
| | BF | GG | RS | MCTS |
| 2mm | 0.2206 | 0.2743 | 0.2716 | **1.4684** |
| *3mm* | 0.1619 | 0.1972 | 0.2044 | 0.3559 |
| *adi* | 0.1857 | 0.1799 | 0.1955 | 0.2071 |
| atax | 2.0037 | 3.3488 | 1.9356 | **3.5023** |
| bicg | 2.4986 | **2.5116** | 1.8525 | 1.8030 |
| covariance | 0.1668 | 0.2614 | 2.0257 | **2.5567** |
| *doitgen* | 0.1165 | 0.1163 | 0.8353 | 0.8135 |
| durbin | 1.0185 | **1.0211** | 1.0205 | 1.0135 |
| *fdtd-2d* | 0.6532 | 0.6534 | 0.6411 | 0.6503 |
| floyd-warshall | 2.7333 | 8.5391 | 8.1178 | **8.5425** |
| gemm | 0.7922 | 1.1433 | 0.9471 | **6.9786** |
| gemver | 0.6213 | 0.8255 | 0.6578 | **1.0221** |
| gesummv | 1.6231 | 1.6391 | 1.5628 | **3.2101** |
| jacobi-1d | 3.5876 | **3.7376** | 3.4982 | 3.6886 |
| jacobi-2d | **2.7010** | 2.6982 | 2.6490 | 2.6906 |
| *lu* | 0.4192 | 0.4248 | 0.4109 | 0.4202 |
| ludcmp | 0.9431 | **1.1567** | 1.1285 | 1.1534 |
| mvt | 0.8202 | 1.7041 | 1.1722 | **1.7395** |
| nussinov | 1.2574 | 1.2548 | **1.2577** | 1.2538 |
| *seidel-2d* | 0.8226 | 0.8221 | 0.8217 | 0.8225 |
| *symm* | 0.7379 | 0.7381 | 0.9046 | 0.7289 |
| syr2k | 0.2855 | 0.3019 | 0.3198 | **2.8777** |
| syrk | 0.4200 | 0.4227 | 0.4097 | **3.4533** |
| *trmm* | 0.1622 | 0.1827 | 0.1702 | 0.1647 |

Figure 4 shows the speedup (with respect to O3P) of the best configuration found over the number of evaluations for different search methods. We observe that the customized MCTS finds high-performing configurations within few iterations (100 to 200) when compared with other search methods. Moreover, because of the restart search strategy, the MCTS is not trapped in a local solution as other search algorithms are (see gemm, 2mm, and covariance).

We investigated why Polly's heuristic was performing well on the eight kernels. We found out that there are transformations that we currently cannot do with directives but that Polly can by using its heuristics. These include unroll-and-jam, loop fusion, loop peeling except for peel(rectangular), wavefronting/skewing, index set splitting, and code motion. The Polly heuristics can also add annotations that help the LoopVectorizer do a better loop optimization.

Figure 5 shows depths of the tree at which the best configuration was found for the search methods. The MCTS finds the best configuration mostly at depth 2 or 3, while other search algorithms find depth 1 or 2. Regarding directives of the best configuration, #pragma clang loop parallelize_thread is included in the best configuration in most cases.

**Table 2: Performance comparison of three ECP proxy applications**

| Applications | Speedup over O3P | | | |
|---|---|---|---|---|
| | BF | GG | RS | MCTS |
| CoMD | 1.3868 | 1.3854 | 1.3849 | **1.3883** |
| miniAMR | 1.0331 | 1.0402 | 0.9970 | **8.4163** |
| SW4lite | 1.1946 | 1.1090 | 1.1094 | **1.2038** |

Table 2 shows a performance comparison of autotuning results on ECP proxy applications. Similar to PolyBench results, we find that our MCTS performs better than O3P, BF, GG, and RS, obtaining the highest speedup over O3P, 1.38, 8.41, and 1.20 for CoMD, miniAMR, and SW4lite, respectively.

To analyze the effectiveness of the search methods, we gathered speedups from all the algorithms and computed a cutoff based on the top 5% speedup values. We then counted the number of configurations that are above this cutoff value over the number of evaluations. Figure 6 shows the results. Besides obtaining the highest speedup configuration, MCTS finds a larger number of configurations above the cutoff value. Moreover, upon further inspection of the best configurations, we found that MCTS finds configurations that contain different pragmas for each application, namely, unrolling and array packing for CoMD, thread parallelization for miniAMR, and tiling with different size selections for SW4lite.

### 4.2 Analysis

We select the gemm kernel for further analysis. First, we employ a decision tree regression method [9] to model the relationship between the pragmas and the observed speedups. We use the resulting decision tree to understand the most important pragmas for the gemm kernel. Specifically, the pragmas that contribute to maximum variance in the speedups are placed at the top of the tree. The decision tree is implemented and visualized by using a Python library, dtreeviz.[5] Figure 7 shows a decision tree, where we can observe that the most important pragma is #pragma clang loop parallelize_thread. Note that given a pragma in the tree, the right (left) of the tree means presence (absence) of the pragma. In addition, the speedup with this pragma is higher in the leaf nodes in the right part of the decision tree than in the left. In the following level of the decision tree, pragmas of different tile sizes are selected as important pragmas, while #pragma clang loop parallelize_thread and #pragma clang loop tile_size(256) emerge as important pragmas for several restarts.

Figure 8 shows the trajectory of the depth explored and the speedup obtained over the number of evaluations. We observe that the search has repetitive phases due to restart. Each phase is characterized by the random walk and the MCTS run. The random walk results in configurations with an equal number of depths. The MCTS exploits the promising configuration found in the random walk by sampling more configurations nearby. The search reaches a convergence, and a restart is performed. The effect of restart is also visible in the second phase. While the search stagnates with a speedup of 13x over the unoptimized code, restart helps MCTS find configurations with a speedup of 14x in the second phase.

---
[5]https://github.com/parrt/dtreeviz

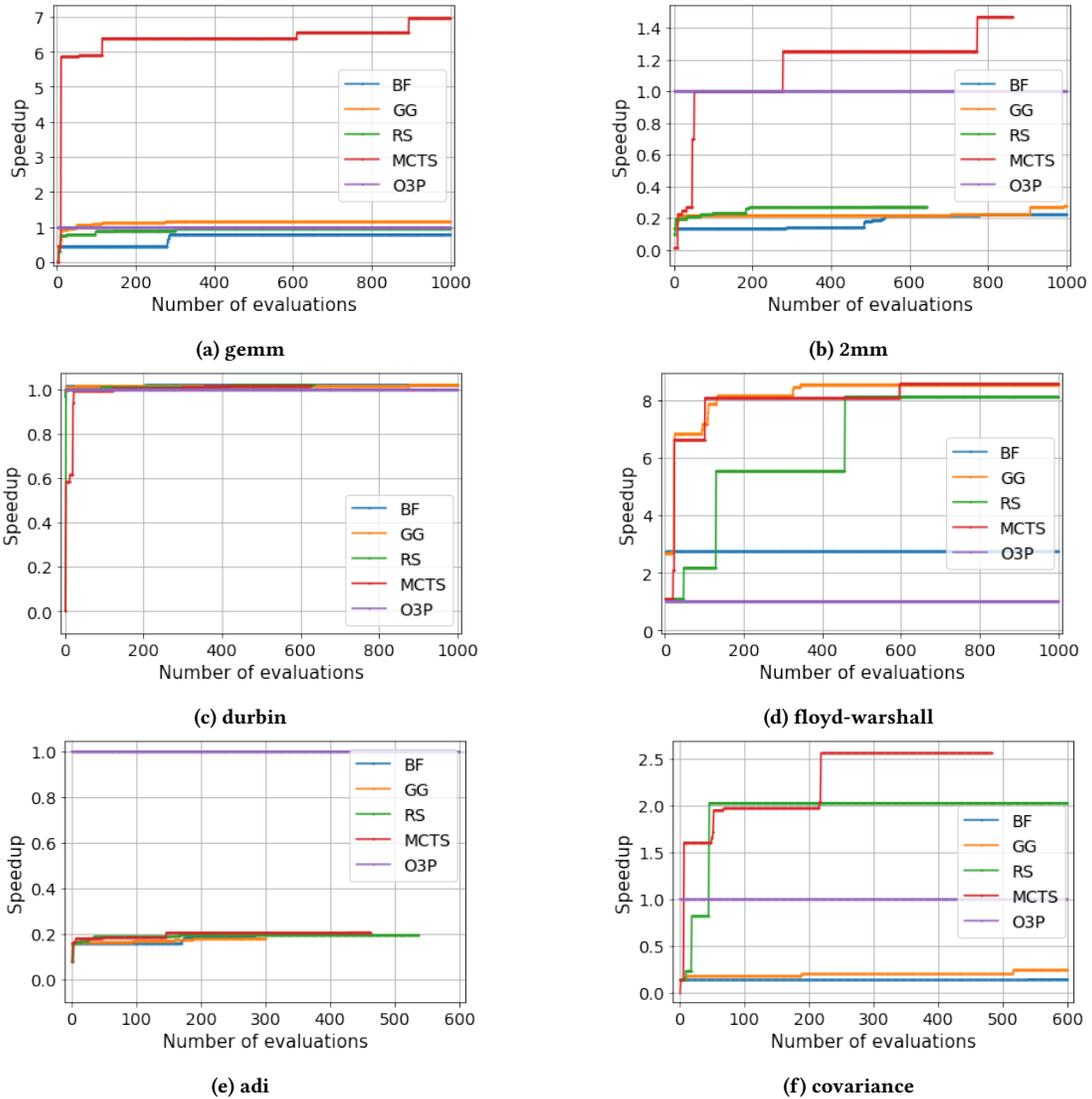

Figure 4: Speedup of the best configuration over number of evaluations for different search methods on PolyBench kernels

We benchmarked Intel's Math Kernel Library (MKL) dgemm routine. It was implemented on the same machine as we used in our other experiments. The execution times obtained by MKL, MCTS, and O3P are 0.0056, 0.0272, and 0.1897, respectively. Even though MCTS outperforms O3P on gemm as well as other kernels, MKL's optimization obtains the fastest time for gemm since it is designed to achieve the highest performance on the Intel machine.

## 5 RELATED WORK

Several autotuning frameworks have been developed to support autotuning regardless of the domain, such as cTuning-cc/Milepost GCC [19], OpenTuner [3], ATF [38], ytopt [5], and mctree [31]. The autotuning frameworks described in [17, 39, 42] are developed for a user-defined sequence of transformations in which only a subset of transformation parameters can be autotuned. There are also domain-specific autotuning frameworks for image and array processing, auch as HalideTuner [48], ProTuner [26], and the works in [1, 35]; AutoTVM [12] and ReLeASE [2] for ML model compilation and accelerators; and PATUS [13] and LIFT [24] for stencil computation.

Many approaches in the literature for solving loop optimization problems use ML and reinforcement learning (RL) and, recently, deep learning approaches. Early studies applied ML approaches to solve a code optimization problem. For example, Stock et al. [41]

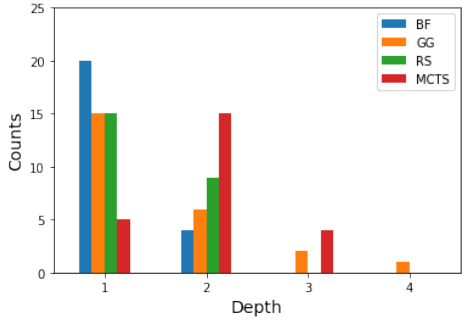

Figure 5: Depth of the best configuration found by different search methods on the PolyBench kernels

applied several ML models such as neural networks, binary trees, SVM, and linear regression with hand-crafted features to solve auto-vectorization for loop permutation, vectorized loop, and unroll. Killian et al. [27] proposed a heuristics algorithm for auto-vectorization based on SVM that provides optimized codes without expert knowledge. The authors investigated how to combine vectorization reports with iterative compilation and code generation to craft features. In their paper, SVM is used to predict the speedup of a program given a sequence of optimization steps.

Recently, state-of-the-art deep learning models are being introduced for loop optimization. In [15], Cummins et al. solved a loop optimization problem by a deep learning model developed based on recurrent neural networks with Long short-term memory. The proposed model takes codes as texts for input and predicts heterogeneous mapping for either GPUs or CPUs and the size of thread coarsening factor $\in \{1, 2, \ldots, 32\}$. They verified their model by benchmarking the previous method based on decision tree [21] on several datasets such as NPB, PolyBench, Parboil, SHOC, and Rodinia. In [2], Ahn et al. developed an RL-based algorithm to solve a compilation problem for deep neural networks built on top of TVM [11]. Haj-Ali et al. [25] proposed a deep RL-based method for loop optimization that determines vectorized factors. Their DNN policy agent takes code embeddings as input states where the embeddings are computed by Code2Vec [43]. The agent determines two vectorized factors such as vectorization width and interleave counts, and run time is used for a reward. The researchers evaluated the proposed algorithm by benchmarking Polly, decision tree, feed-forward networks, and nearest-neighbor search. In [46], an autotuning framework based on Bayesian optimization (BO) was proposed to explore the parameter space search. BO models such as random forests, extra trees, gradient-boosted regression trees, and Gaussian processes are selected as a learner in the proposed framework. Performance of the framework is tested on six PolyBench kernels with the latest LLVM Clang/Polly loop optimization pragmas. However, these frameworks are limited to the search space expressed in a vector space. A search tree considered in the literature does not directly map to the space of a fixed number of knobs.

Recent literature solves loop optimization by defining search space as a tree, such as in [1] and [26] for the Halide scheduler, in [7] for Telamon, and in [31] for LLVM Clang/Polly loop optimization pragmas.

Tree search algorithms including MCTS are introduced to optimize Halide [36] schedules. To tackle the limitation of a vector search space for Halide studied in [48] and [35], Adams et al. [1] proposed an autotuning framework based on beam search over a tree-shaped search space. Since each node in the search tree represents an intermediate schedule, a cost model based on neural networks that output the dot product of a vector of hand-designed terms and a vector of coefficients is used to predict the run time. The authors verified that the proposed framework finds better configurations than the previous method, HalideTuner [48], does. Haj-Ali et al. [26] further improved Halide schedule autotuning by applying MCTS, which tackles the non-global behavior of the previous beam search approach. The authors report that the schedules found by MCTS are up to 3.6 times faster than those found by existing search methods such as random search, greedy search with $k$ of 1, and the previous beam search. Since their problem space is only Halide schedules, however, these methods do not apply to arbitrary source C/C++ source codes as our schedules do.

## 6 CONCLUSIONS AND FUTURE WORK

We developed an autotuning framework to identify high-performing pragma combinations for loop transformations. Based on LLVM/Clang and Polly, our search space has a tree structure and grows dynamically such that the space is potentially infinite. To efficiently explore a large search space, we developed a customized Monte Carlo tree search (MCTS) into the autotuning framework together with a new reward mechanism, restart strategy, and transfer learning to improve the search efficiency. We evaluated the proposed framework on 24 PolyBench kernels and three ECP proxy applications. In the experiments, the proposed MCTS surpasses Polly's heuristic optimization on 16 PolyBench kernels and all three ECP proxy applications. In addition, our method outperforms other search algorithms for breadth-first, global greedy, and random search. We obtain a speedup over Polly's optimization of 2.30 on average in the whole experiment, compared with the other methods that achieve 1.06, 1.40, and 1.35 for the breadth-first, global greedy, and random search, respectively.

For our future work, we plan to further refine the tree search algorithm as well as the search space. We plan to implement the remaining loop transformations that the Polly heuristic can use (unroll-and-jam, loop distribution/fusion, etc.) but are currently not part of the search space. With these implemented, we expect that autotuning will be better than Polly's compile-time heuristic. We currently do not prune the search space for infeasible transformations (such as reversing a loop twice) and identical configurations that can be reached by different transformation sequences. The latter would make the search space a directed acyclic graph instead of a tree.

## ACKNOWLEDGMENT

This research was supported by the Exascale Computing Project (17-SC-20-SC), a collaborative effort of the U.S. Department of Energy Office of Science and the National Nuclear Security Administration.

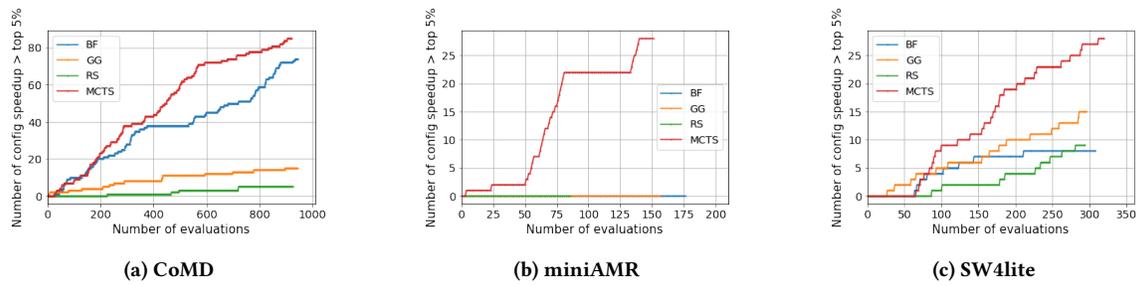

Figure 6: Number of configurations that are above this cutoff value over the number of evaluations

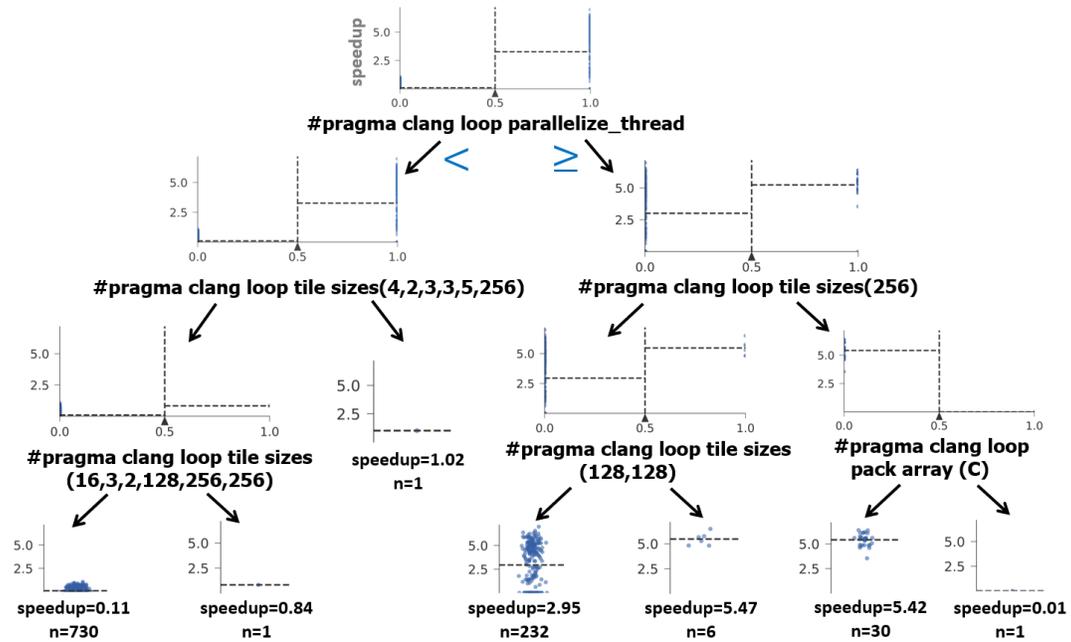

Figure 7: Decision tree visualizing relationships between pragmas and observed speedup for the gemm kernel

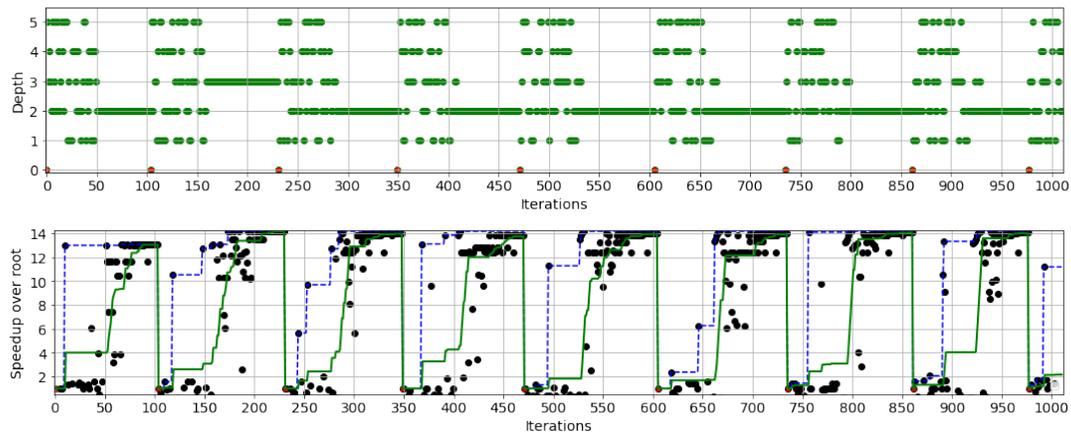

Figure 8: Trajectory of depth and speedup by the MCTS algorithm. Green and blue lines show the target value and the best speedup obtained over the iterations.